\documentclass[journal,comsoc]{IEEEtran}

\usepackage{amsmath,amsfonts,amssymb,amsthm,epsfig,epstopdf,titling,url,array}

\usepackage[colorlinks]{hyperref} 
\usepackage[table]{xcolor}

\usepackage[utf8]{inputenc}
\usepackage[english]{babel}
 
\usepackage{amsthm}
 \newtheorem{theorem}{Theorem}
 
\theoremstyle{definition}
\newtheorem{example}{Example}

\usepackage{amsmath}

\begin{document}

\title{Polar codes for secret sharing}

\author{Mohsen Moradi
\thanks{M. Moradi was with the Department of Mathematics, Amirkabir University of Technology - Tehran Polytechnic , Tehran, Iran, ( e-mail:
m.moradi8111@aut.ac.ir ).}}

\maketitle

\begin{abstract}
A secret can be an encrypted message or a private key to decrypt the ciphertext. One of the main issues in cryptography is keeping this secret safe. Entrusting secret to one person or saving it in a computer can conclude betrayal of the person or destruction of that device. For solving this issue, secret sharing can be used between some individuals which a coalition of a specific number of them can only get access to the secret.  In practical issues, some of the members have more power and by a coalition of fewer of them, they should know about the secret. In a bank, for example, president and deputy can have a union with two members by each other. In this paper, by using Polar codes secret sharing has been studied and a secret sharing scheme based on Polar codes has been introduced. Information needed for any member would be sent by the channel which Polar codes are constructed by it. 
\end{abstract}


\IEEEpeerreviewmaketitle

\section{Introduction}

\IEEEPARstart{I}{N} communication, in most of the cases, secrecy of information is the most important issue. If a person wants to send an information, he can encrypt it, or by using steganography the information can be hidden in others view \cite{moradi2017combining}. The public key encryption is one of the main concepts in cryptography which has been introduced by Diffie and Hellman, and it was an answer to key exchanging old problem and refers to the digital signature \cite{diffie1976new}. In mathematics, public key cryptography is known as the one-way function which uses trapdoor function. For securing an information we can encrypt it, but for protecting encryption key other ways should be tried. One way for protecting key is keeping it safe to a special place (in a computer, in human memory, or in another safe place). This method is unreliable in that by only one misfortune (the corruption of computer, sudden death of the person, or intentional sabotage) the key can be lost. One simple solution is saving some copies of the key to different places, but this is also insecure as well (infiltration in computer, betrayal, or human error). The solution which can answer to key security or any other secrecy is sharing that secret among some individuals that only a coalition of a specific number of them can rebuild the secret. Different kinds of secret sharing schemes have been introduced which in this paper briefly they would be described and a new secret sharing scheme based on Polar codes would be introduced. In section \ref{section 2} a review of different secret sharing schemes has been mentioned and in section \ref{section 3} construction of Polar codes which is suitable for secret sharing is studied. In section \ref{section 4} Polar code based secret sharing is introduced and there are two examples to illustrate it.

\section{	Secret sharing scheme }\label{section 2}

Assume that a group of 11 scientists wants to securely preserve their secret information. They can easily settle the information in a strongbox. Now if they want to install some locks on strongbox which only by a coalition of six of them they can open its door, in a combinatorial way the minimum number of locks and the necessary amount of keys for any scientists can be determined. This strongbox needs $(11, 6)=462$ locks and any scientist should hold $(10, 5)=252$ keys with him \cite{Liu}. Now if this secret should be shared among individuals which are a lot more than 11 people, answer to this issue needs too many locks and keys which its amount would increase exponentially. 

The first and easiest answer to this problem is using only one lock and giving the key to one person. In confidentiality aspect, this idea is not good at all. The person can lose the key, forget the secret, or even he can betray. From the time that electronic communication has been common in human life, the value of safe communication has been increased. The notion of information security for different needs and different purposes may vary. Anyway, without considering the way which we use, secrecy is dependent to the key confidentiality which only legal members should be aware of it. 

The necessity of key confidentiality has some arguments like establishing key secrecy on only one member or one database can decrease system's secrecy level to the person or database's reputation and if other copies of key do not be available, it can be lost, or the software or hardware can be damaged. On the other hand, if the key is held by more than one member, an enemy would have more flexibility in achieving to the key which can decrease its security. 

Shamir \cite{shamir1979share} generalized this problem in a way that for a secret $s$, the goal is dividing $s$ to $N$ shares like $s_{1}, s_{2}, \cdots , s_{N}$  which only by coalition of $t$ or more than this $s_{i}'s$ the secret be computable and by knowing any of the $t-1$ or fewer shares it would not be possible to understand anything about secret $s$. That means all the possible answers for $s$ have the same probability. This kind of scheme is known as a $(t,n)$ threshold scheme. Shamir had used polynomial interpolation for answering secret sharing scheme. Another idea is based on hyperplanes which Blakley introduced it \cite{blakley1979safeguarding}. In Asmuth-Bloom secret sharing scheme the secret is shared by a modular arithmetic and secret would be reassembled by Chinese remainder theorem. Mignotte secret sharing which has some integer sequences known as Mignotte sequences is based on Chinese remainder theorem \cite{mignotte1983share}. Like secret sharing scheme, function sharing scheme needs to distribute function computation based on function sharing scheme in a way that each part of the calculation is done by different members. The partial results of computations should be combined with each other until the final output value of the function, without revealing partners secret, (the partial calculations made by members) be done \cite{de1994share}. Encryption functions and digital signature can be used as examples for sharing functions by using secret sharing scheme for public key algorithms like RSA and ElGamal \cite{elgamal1985public}.  Also, the security of new devices like electronic voting devices is dependent on secret sharing functions. RSA secret system is the most used system in public key secret system \cite{mao2003modern}. RSA signature and encryption functions can be shared among members by Asmuth-Bloom secret sharing system \cite{asmuth1983modular}. RSA threshold signature scheme is secure, that it means the scheme is non-counterfeiting in an adaptive chosen-plaintext attack, and in the result, RSA function is trapdoor \cite{mao2003modern}. In this scheme it has been assumed that adversary is static, it means he can control $t-1$ members. Thsese members can be corrupted individuals and adversary even can request the signature of messages from them but he cannot do anything about other members which means adversary is static. 

Digital Signature Standard (DSS) is a criterion for digital signature in the USA and DSS sharing is an interesting problem which a solution has been introduced by Gennaro and et al. based on Shamir secret sharing scheme \cite{gennaro1996robust}. Asmuth-Bloom secret sharing scheme is probably the first DSS threshold signature scheme based on Chinese remainder theorem. DSS threshold signature scheme is secure which means is non-counterfeiting in adaptive chosen plain-text attack \cite{kaya2014sharing}.

Also, linear codes have been used for secret sharing. The relation between minimal codewords of dual codes and secret sharing has been studied \cite{massey1993minimal}. Another great researches in secret sharing by linear codes area also has been done \cite{bertilsson1993construction}. Arikan in his paper introduced channel capacity achieving Polar codes which in following a secret sharing scheme based on it would be constructed.

\section{Construction of suitable Polar codes for secret sharing}\label{section 3}

Polar codes introduced by Arikan has a construction which is based on the channel it uses. At first Polar codes have been used for Binary input Discrete Memoryless Symmetric Channels (B-DMSC) and it is shown these codes can achieve to the channel capacity by $O(N\log{N})$ encoding and decoding complexity. $N$ is a power of 2 and is code length \cite{arikan2009channe}. For showing a B-DMSC a $W:\mathcal{X} \longrightarrow \mathcal{Y}$ channel by input alphabet $\mathcal{X}$ and output alphabet $\mathcal{Y}$ with transition probability function $W(y\mid x)$  which $y \in \mathcal{Y}$ and $x \in \mathcal{X}$ has been used. In this paper binary input alphabet, $\mathcal{X}=\lbrace 0 , 1 \rbrace $ is assumed. For showing a channel which has been used by $N$ times using of $W$, $W^{N}:\mathcal{X}^{N}\longrightarrow \mathcal{Y}^{N}$ notation has been used which $ W^{N} (y_{1}^{N}\mid x_{1}^{N} )=\prod_{i=1}^{N} W(y_{i} |x_{i})$. For a symmetric channel $W$, the capacity is as 
 
 \begin{equation}\label{eq rate}
 I(W)=\sum_{x\in \mathcal{X}} 
\sum_{y\in \mathcal{Y}}
\dfrac{1}{2}
W(y \mid x) 
\log{           \dfrac{        W(y \mid x)         }     {        \dfrac{1}{2}W(y \mid 0 ) + \dfrac{1}{2}W(y \mid 1)            }              },
 \end{equation}
 
And the Bhattacharyya parameter is defined as

\begin{equation} \label{Bhattacharyya}
Z(W)= \sum_{y\in \mathcal{Y}} \sqrt{W(y \mid 0)W(y \mid 1)}.
\end{equation}

$Z(W)$ is an approximation for the reliability of the channel. If all inputs of the channel $W$ assumed to have the same probability, the value of $I(W)$ is the highest rate and $Z(W)$  is an upper bound for maximum-likelihood error decision which $W$ has been used for only one time for a 0 or 1 to the channel. The logarithm base is 2 and both $I(W)$ and $Z(W)$ parameter's values are in [0,1] interval. It is proved that $I(W)\simeq 1$ if and only if $ Z(W) \simeq 0$ \cite{arikan2009channe}. In channel polarization by using $N$ independent copies of B-DMSC $W$, $N$ $ \lbrace W_{N}^{(i)}: 1 \leq i \leq N \rbrace $ polarized channels would be obtained. As $N$ increases, the capacity of these symmetric channels, except a few fractions of them, tends to 0 or 1 and channel polarizing is done in two steps: channel combining and channel splitting. 

In channel polarizing, for achieving the symmetric channel capacity $I(W)$, each $I(W_{N}^{(i) })$  is seen separately and those channels which $Z(W_{N}^{(i) }) $ for them is near to zero would be chosen. Set of $i$'s indices which $Z(W_{N}^{(i) }) $ for them is near to zero is called information or unfrozen bits and this set is shown by $\mathcal{A}$. The remaining indices are called frozen bits and is shown by $\mathcal{A}^{c}$ which $Z(W_{N}^{(i) }) $ for them is near to 1. In the construction of Polar codes, a class of block codes which contains Polar codes are used. Consider $G_{N}$ matrix which is $n$ times Kronecker product of  matrix
$G=\begin{bmatrix}
1 & 0 \\
1 & 1
\end{bmatrix}$.
The length of constructed codes in this way is a power of 2-i.e., $N=2^{n}$. For a given $N$, any message in this class is encoded in the same way: $x_{1}^{N}=u_{1}^{N} G_{N}$. For the set of information bits $\mathcal{A}$ in $\lbrace 1, \cdots ,N\rbrace $ the codeword is:

\begin{equation}
x_{1}^{N}=u_{A} G_{N} (\mathcal{A})\oplus u_{A^{c} } G_{N} ({A^{c}}) .
\end{equation}

$G_{N} (\mathcal{A})$'s row space is a subspace of $G_{N}$'s row space which is determined by unfrozen bits. If $\mathcal{A}$ and $u_{A^{c} }$ considered being fixed ( it can be assumed that $u_{\mathcal{A}^{c}}$  is zero) and $u_{\mathcal{A}}$ considered to be an arbitrary vector, a map from $u_{\mathcal{A}}$ to the $x_{1}^{N}$ can be obtained. This is a cosset coding: a cosset of linear block codes by generator matrix $G_{N} (\mathcal{A})$ which every cosset is determined by a fixed vector $u_{\mathcal{A}^{c}}  G_{N} (\mathcal{A}^{c})$. This set is called cosset codes of $G_{N}$ and it is shown by $(N,K,\mathcal{A},u_{\mathcal{A}^{c} })$ parameters which $K$ is the dimension of the code and determines the size of $\mathcal{A}$. $u_{\mathcal{A}^c }$ is also called fixed bits. Polar codes can be constructed by determining $\mathcal{A}$ information bits. 
$H_{N}$, the dual code of $G_{N}$, is some rows of a matrix which can be obtained from $n$ times Kronecker product of matrix 
$H=\begin{bmatrix}
1 & 1 \\
0 & 1
\end{bmatrix}$. The rows are chosen based on frozen bits in  $\mathcal{A}^{c}$. It can easily be seen that $h_{i,j}=g_{N-1-i,N-1-j}$. If a row of Polar codes is chosen based on information bit of $C$, that row of $H_{N}$ is correspondent to a frozen bit of its dual code and vice versa. 
For matrix $G$, a permutation $\pi$ on its rows can be defined in a way that
 $G=\pi (G)=
 \begin{bmatrix}
 G_{F}\\
 G_U
 \end{bmatrix} $
 which $G_{F}$ is corresponding to frozen bits and $G_{U}$ is corresponding to the unfrozen bits. In conclusion, 
$ G^{T}=
\begin{bmatrix}
H_U\\
H_F 
 \end{bmatrix}  $
in a way that  $G_{U} H_{U}^{T}=0$. In result, a systematic encoding for solving $[x_{F} x_{U} ] H_{U}^{T}=0$ is used that $x_{U}$ is for information bits. Notice that permutation $\pi$ for obtaining
$G=\pi (G)=
 \begin{bmatrix}
 G_{F}\\
 G_U
 \end{bmatrix} $
from the initial matrix $G$ is necessary to be done to obtain $[x_{F} x_{U} ]$.
\begin{example}
$$H_{8}=G_{8}^{T}  ; $$
$G_{U}=G_{8} ([4 \ 6 \ 7 \ 8]),:)$ ;\\
$H_{U}=H_{8} ([1 \ 2 \ 3 \ 5], :)$ ;
$$ \Longrightarrow  mod((G_{U} \times H_{U}^{T} ),2)=0. $$ 
$$H_{16}=G_{16}^{T}  ; $$
$G_{U}=G_{16} ([8 \ 10 \ 11 \ 12 \ 13 \ 14 \ 15 \ 16],:)$ ;\\
$H_{U}=H_{16} ([1 \ 2  \ 3 \ 4 \ 5 \ 6 \ 7 \ 9], :)$; 
$$ \Longrightarrow mod( (G_{U} \times H_{U}^{T} ),2 )=0.    $$
\end{example}
In Polar code generator matrix, its row weight can easily be calculated \cite{li2015error}:
for $N=2^{n}$, every column of matrix $G_{N}$ has a weight equal to
 $2^{w_{H} ( \overline{b}_{1}^{j},  \overline{b}_{2}^{j},\cdots , \overline{b}_{n}^{j})}  $
 which $(b_{1}^{j},b_{2}^{j},\cdots , b_{n}^{j})$ is binary expansion of $j-1$ and $\overline{b}_{i}^{j}=b_{i}^{j}\oplus 1$.
For example, weights of some codewords are as table \ref{table:1}. 

\renewcommand{\arraystretch}{1.7}
\begin{table}[h!]
\small
\caption{column weights of $G_{8}$}
\begin{center}
\begin{tabular}{ | c | | c | | c | | c | | c |}

 \hline 
 \multicolumn{1}{|c||}{column j}     &  \multicolumn{1}{|c||}{j-1}     &  \multicolumn{1}{|c||}{$(b_{1}^{j},b_{2}^{j}, b_{3}^{j})$}   &  \multicolumn{1}{|c||}{$ ( \overline{b}_{1}^{j},  \overline{b}_{2}^{j} , \overline{b}_{3}^{j})$}  &  \multicolumn{1}{|c||}{ weight}         \\
 \hline \hline
1   & \multicolumn{1}{|c||}{0}   & \multicolumn{1}{|c||}{(0 0 0)}  & \multicolumn{1}{|c||}{(1 1 1)} & \multicolumn{1}{|c||}{8}    \\
 \hline
2   & \multicolumn{1}{|c||}{1}   & \multicolumn{1}{|c||}{(0 0 1)}  & \multicolumn{1}{|c||}{(1 1 0)} & \multicolumn{1}{|c||}{4}    \\
 \hline
3   & \multicolumn{1}{|c||}{2}   & \multicolumn{1}{|c||}{(0 1 0)}  & \multicolumn{1}{|c||}{(1 0 1)} & \multicolumn{1}{|c||}{4}    \\
 \hline
4   & \multicolumn{1}{|c||}{3}   & \multicolumn{1}{|c||}{(0 1 1)}  & \multicolumn{1}{|c||}{(1 0 0)} & \multicolumn{1}{|c||}{2}    \\
 \hline
5   & \multicolumn{1}{|c||}{4}   & \multicolumn{1}{|c||}{(1 0 0)}  & \multicolumn{1}{|c||}{(0 1 1)} & \multicolumn{1}{|c||}{4}    \\
 \hline
6   & \multicolumn{1}{|c||}{5}   & \multicolumn{1}{|c||}{(1 0 1)}  & \multicolumn{1}{|c||}{(0 1 0)} & \multicolumn{1}{|c||}{2}    \\
 \hline
7   & \multicolumn{1}{|c||}{6}   & \multicolumn{1}{|c||}{(1 1 0)}  & \multicolumn{1}{|c||}{(0 0 1)} & \multicolumn{1}{|c||}{2}    \\
 \hline
8   & \multicolumn{1}{|c||}{7}   & \multicolumn{1}{|c||}{(1 1 1)}  & \multicolumn{1}{|c||}{(0 0 0)} & \multicolumn{1}{|c||}{1}    \\

 \hline
\end{tabular}
\end{center}
\label{table:1}
\end{table}

\section{Polar code based secret sharing}\label{section 4}

In secret sharing, binary codes can be as goos as non-binary codes. Notice that any secret space can be encoded to other secret space which every secret is a binary stream by finite length. 

The length of any binary stream can be fixed or be variable. For this reason, for using Polar codes, it is assumed that secret space is a binary stream. Because of that, the secret can be shared among a group of members as a bit and reconstructed bit by bit by the share of any member. In result, it is assumed that secret space is $\lbrace 0,1\rbrace$. For simplicity, in this paper secret can take 0 and 1 values by same probability. 
Consider a binary $(N,k)$ Polar code which $N=2^{n}$ is the code length and code is obtained by $n$ times using of $W$ channel. $k$ is the size of unfrozen set and $G=(g_{1},\cdots ,g_{N} )$ is the generator matrix of Polar code. Fix index $p$ in $\mathcal{A}$ which would be used in constructing secret from Polar code. We have secret $s\in \lbrace 0,1 \rbrace$ and  $P_{1},\cdots ,P_{w}$ members that $w\leq k$. For secret sharing, a random vector $u=(u_{1},\cdots ,u_{k} )\in F_{2}^{k}$ is chosen in a way that $s=ug_{p}$. Dealer uses $u$ as information vector and would compute codeword $t=(t_{1},\cdots ,t_{N})=uG$. The dealer gives $t_{i}$ to the $P_{i}$ member which $i\in \mathcal{A}$ and $i\neq p$. It can be calculated that $t_{p}=ug_{p}=s$. A set of $\lbrace t_{i_{1}}, t_{i_{2}},\cdots ,t_{i_{m}} \rbrace$ shares would determine secret $s$ if and only if $g_{p}$ be a combination of $g_{i_{1}}, \cdots ,g_{i_{m}}$ values. In other words, a set of $ t_{i_{1}}, t_{i_{2}},\cdots ,t_{i_{m}}$ shares can determine the value of secret if and only if there is a codeword $(0,0,…,0,c_{i_{1}} , 0, \cdots ,1,\cdots ,c_{i_{m}} , 0,\cdots ,0)$ in dual code of polar code which its $p$ element is 1. In result, $g_{p}=\sum_{j=1}^{m} x_{j} g_{i_{j}}$ is obtained and $x_{j}\in F_{2}$ and $s=\sum_{j=1}^{m} x_{j} t_{i_{j}} $ . 

A group of members is called minimal access set if they can compute secret by their shares and any proper subset of it would not be able to find access to the secret. Support of a $c\in F_{2}^{N}$ vector is defined as $\lbrace 1\leq i \leq N : c_{i}\neq  0\rbrace$. A codeword $c_{2}$ covers codeword $c_{1}$ if support of $c_{2}$ contains support of $c_{1}$. A codeword is called minimal if it covers only its coefficients and does not cover any other nonzero vector. If $p$'s element of a codeword is 1, it is called minimal $p$-codeword. 

\begin{theorem}
Suppose $C$ is an $(N,k)$ Polar code by $G=(g_{1},\cdots ,g_{N})$ as its generator matrix which any nonzero codeword of $C$ is a minimal vector. In secret sharing based on the dual code, totally there are $2^{k-1}$ minimal access sets and if the $g_{i}$ is a multiplication of $g_{p}$ for  $1\leq i\leq N$ which $i\neq p$, member $P_{i}$ should be in any minimal access set. This member is called a dictator member. Also, if the $g_{i}$ is a multiplication of $g_{p}$ for  $1\leq i\leq N$ which $i\neq p$, $P_{i}$ member should be in $2^{k-2}$ from $2^{k-1}$ minimal elements sets. 

\end{theorem}

A more general proof of this theorem there is in \cite{ding2003covering}. Based on this theorem, it is better to construct codes which any nonzero codeword be a minimal vector. If weights of a linear code are close to each other, from the next theorem it can be concluded that every nonzero codeword is minimal.

\begin{theorem}
In $(N,k)$ code $C$, suppose that $w_{min}$ and $w_{max}$ are minimum and maximum weights, respectively. If $\dfrac{w_{min}}{w_{max}} > \dfrac{1}{2}$, every nonzero codeword of $C$ is minimal. 
\end{theorem}
Based on this theorem \cite{ding2003covering}, those rows of $G_{N}$ can be chosen which have weights near to each other:

\begin{center}
$G_{8} = $
\renewcommand\arraystretch{1} 
$\begin{bmatrix}
1 & 0 & 0 & 0 & 0 & 0 & 0 & 0 \\
1 & 1 & 0 & 0 & 0 & 0 & 0 & 0 \\
1 & 0 & 1 & 0 & 0 & 0 & 0 & 0 \\
\rowcolor{green} 1 & 1 & 1 & 1 & 0 & 0 & 0 & 0 \\
1 & 0 & 0 & 0 & 1 & 0 & 0 & 0 \\
\rowcolor{green} 1 & 1 & 0 & 0 & 1 & 1 & 0 & 0 \\
\rowcolor{green} 1 & 0 & 1 & 0 & 1 & 0 & 1 & 0 \\
1 & 1 & 1 & 1 & 1 & 1 & 1 & 1 
\end{bmatrix}$
\end{center}

The result code is as: $$x=[0 0 0 1 0 1 1 0] G_{8}.$$

In the following two examples of secret sharing scheme constructed by $(8, 4)$ and $(32, 16)$ Polar codes which secret is computed by $g_{p}$ are given. A coalition of each one for rebuilding secret is introduced. 

\begin{example}
We are looking for a secret sharing scheme based on a Polar code by a length of 8 and dimension of 4. Suppose the channel is BEC(1/2). Values of $Z(W_{N}^{(i) }) $s are as follows: 

0.9961, 0.8789, 0.8086, \textcolor{blue}{\textbf{0.3164}}, 0.6836, \textcolor{blue}{\textbf{0.1914}}, \textcolor{blue}{\textbf{0.1211}}, \textcolor{blue}{\textbf{0.0039}}.

 Because $k=4$, four values of $Z(W_{N}^{(i) }) $s which is fewer than others are specified by blue color and those are in positions 4, 6, 7, 8.  So $\mathcal{A}=\lbrace 4,6,7,8\rbrace $ is the information set and $G_{U}$ matrix is defined based on it. $G_{U}$ and $H_{U}$ matrices by their row's weights are shown in tables \ref{table:2} and \ref{table:3}. 

\renewcommand{\arraystretch}{1.7}
\begin{table}[h!]
\small
\caption{$G_{U}$ matrix and its row weights}
\begin{center}
\begin{tabular}{ | c |  | c |}

 \hline 
 \multicolumn{1}{|c||}{$G_{U}$ rows}     &  \multicolumn{1}{|c||}{row weights}     \\
 \hline \hline
 \multicolumn{1}{|c||}{[1 1 1 1 0 0 0 0]}   & \multicolumn{1}{|c||}{4}      \\
 \hline
  \multicolumn{1}{|c||}{[1 1 0 0 1 1 0 0]}   & \multicolumn{1}{|c||}{4}      \\  
   \hline
  \multicolumn{1}{|c||}{[1 0 1 0 1 0 1 0]}   & \multicolumn{1}{|c||}{4}      \\  
   \hline
  \multicolumn{1}{|c||}{[1 1 1 1 1 1 1 1]}   & \multicolumn{1}{|c||}{8}      \\
 \hline

\end{tabular}
\end{center}
\label{table:2}
\end{table}

\begin{table}[h!]
\small
\caption{$H_{U}$ matrix and its row weights}
\begin{center}
\begin{tabular}{ | c |  | c |}

 \hline 
 \multicolumn{1}{|c||}{$H_{U}$ rows}     &  \multicolumn{1}{|c||}{row weights}     \\
 \hline \hline
 \multicolumn{1}{|c||}{[1	1	1	1	1	1	1	1]}   & \multicolumn{1}{|c||}{8}      \\
 \hline
  \multicolumn{1}{|c||}{[0	1	0	1	0	1	0	1]}   & \multicolumn{1}{|c||}{4}      \\  
   \hline
  \multicolumn{1}{|c||}{[0	0	1	1	1	0	1	1]}   & \multicolumn{1}{|c||}{4}      \\  
   \hline
  \multicolumn{1}{|c||}{[0	0	0	0	1	1	1	1]}   & \multicolumn{1}{|c||}{4}      \\
 \hline

\end{tabular}
\end{center}
\label{table:3}
\end{table}

In secret sharing based on this code, index $p$ is fixed. So for a random vector $u=(0, 0, 0, u_{4}, 0, u_{6}, u_{7}, u_{8})$, secret is $s=ug_{p}$. Dealer uses this $u$ as information vector and obtains codeword $t=(t_{1},\cdots ,t_{8})=uG$ and by using BEC, he sends  $t_{1},\cdots  , t_{8}$ shares to the $P_{1},\cdots, P_{8}$ members which $P_{1},P_{2}, P_{3}, P_{4}$ are called imaginary members. If we determine $p=8$, for rebuilding secret a coalition of at least three members is necessary which can be one of the  $\lbrace P_{1},\cdots, P_{7}\rbrace$, $\lbrace P_{2}, P_{4}, P_{6}\rbrace$, $\lbrace P_{3}, P_{4}, P_{7}\rbrace$, or $\lbrace P_{5},P_{6}, P_{7}\rbrace$ coalitions. Because actually received shares for imaginary four $P_{1}, P_{2}, P_{3}, P_{5}$ members are fixed, by a coalition of two members of $\lbrace P_{4}, P_{6}\rbrace$, $\lbrace P_{4}, P_{7}\rbrace$, or  $\lbrace P_{6}, P_{7}\rbrace$ the secret can be calculated. If $p\neq N$, $P_{N}$ would be a dictator member and should be present in any coalition. It can be practical when an administrator should be present in any important meeting. 
\end{example}

\begin{example}
A secret sharing scheme based on Polar codes with length of 32 and dimension of 16 on AWGN channel which $\sigma =0.9$ would be constructed. For this code, values of $Z(W_{N}^{(i) })$ are as:

 1, 0.9999, 0.9998, 0.9623, 0.9987, 0.9315, 0.8914, 0.502, 0.9937, 0.8685, 0.8065, \textcolor{blue}{\textbf{0.3584}}, 0.7134, \textcolor{blue}{\textbf{0.2505}}, \textcolor{blue}{\textbf{0.1803}}, \textcolor{blue}{\textbf{0.01}}, 0.9755, 0.7529, 0.6673, \textcolor{blue}{\textbf{0.2083}}, 0.5559, \textcolor{blue}{\textbf{0.1309}}, \textcolor{blue}{\textbf{0.0878}}, \textcolor{blue}{\textbf{0.0022}}, \textcolor{blue}{\textbf{0.4281}}, \textcolor{blue}{\textbf{0.0704}}, \textcolor{blue}{\textbf{0.0451}}, \textcolor{blue}{\textbf{0.0006}}, \textcolor{blue}{\textbf{0.027}}, \textcolor{blue}{\textbf{0.0002}}, \textcolor{blue}{\textbf{0.0001}}, \textcolor{blue}{\textbf{0}}.

Because k=16, information set is $\mathcal{A}=\lbrace 12, 14, 15, 16, 20, 22, 23, 24, 25, 26, 27, 28, 29, 30, 31, 32\rbrace$. Based on information bits, $G_{U}$ would be constructed. For secret sharing scheme based on this code, if we fix $p=32$, according to random vector
$u=(u_{12}, u_{14}, u_{15}, $
$ u_{16}, u_{20}, u_{22}, u_{23}, u_{24}, u_{25}, u_{26}, u_{27}, 
 u_{28}, u_{29}, u_{30}, u_{31}, u_{32}),$ secret is $s=ug_{p}$. Dealer uses u as information vector and obtains codeword $t=(t_{1},\cdots ,t_{32} )=uG$, and by using binary input AWGN channels, gives $t_{i}$ to the member $P_{i}$ which $i\in \mathcal{A}$ and $i\neq p$. Coalitions based on each row of matrix $H_{U}$ is given in table \ref{table:4}.

\begin{table}[h!]
\small
\caption{$H_{U}$ matrix and its row weights}
\begin{center}
\begin{tabular}{ | c |  | c |}

 \hline 
 \multicolumn{1}{|c||}{row}     &  \multicolumn{1}{|c||}{coalition of members to rebuild secret}     \\
 \hline \hline
 \multicolumn{1}{|c||}{1}   & \multicolumn{1}{|c||}{$P_{12}, P_{14}, P_{15}, P_{16}, P_{20}, P_{22} ,\cdots , P_{31}$}      \\
 \hline
  \multicolumn{1}{|c||}{2}   & \multicolumn{1}{|c||}{$P_{12}, P_{14}, P_{16}, P_{20}, P_{22}, P_{24}, P_{26}, P_{24}, P_{28}, P_{30}$}      \\
 \hline
  \multicolumn{1}{|c||}{3}   & \multicolumn{1}{|c||}{$P_{12}, P_{15}, P_{16}, P_{20}, P_{23}, P_{24}, P_{27}, P_{28}, P_{31}$}      \\
 \hline
  \multicolumn{1}{|c||}{4}   & \multicolumn{1}{|c||}{$P_{12}, P_{16}, P_{20}, P_{24}, P_{28}$}      \\
 \hline
  \multicolumn{1}{|c||}{5}   & \multicolumn{1}{|c||}{$P_{14}, P_{15}, P_{16}, P_{22}, P_{23}, P_{24}, P_{29}, P_{30}, P_{31}$}      \\
 \hline
  \multicolumn{1}{|c||}{6}   & \multicolumn{1}{|c||}{$P_{14}, P_{16}, P_{22}, P_{24}, P_{30}$}      \\
 \hline
  \multicolumn{1}{|c||}{7}   & \multicolumn{1}{|c||}{$P_{15}, P_{16}, P_{23}, P_{24}, P_{31}$}      \\
 \hline
  \multicolumn{1}{|c||}{8}   & \multicolumn{1}{|c||}{$P_{16}, P_{24}$}      \\
 \hline
  \multicolumn{1}{|c||}{9}   & \multicolumn{1}{|c||}{$P_{12}, P_{14}, P_{15}, P_{16}, P_{25}, P_{27}, \cdots, P_{31}$}      \\
 \hline
  \multicolumn{1}{|c||}{10}   & \multicolumn{1}{|c||}{$P_{12}, P_{14}, P_{16}, P_{26}, P_{28}, P_{30}$}      \\
 \hline
  \multicolumn{1}{|c||}{11}   & \multicolumn{1}{|c||}{$P_{12}, P_{15}, P_{16}, P_{27}, P_{28}, P_{31}$}      \\
 \hline
  \multicolumn{1}{|c||}{12}   & \multicolumn{1}{|c||}{$P_{14}, P_{15}, P_{16}, P_{29}, P_{30}, P_{31}$}      \\
 \hline
  \multicolumn{1}{|c||}{13}   & \multicolumn{1}{|c||}{$P_{20}, P_{22}, \cdots, P_{31}$}      \\
 \hline
  \multicolumn{1}{|c||}{14}   & \multicolumn{1}{|c||}{$P_{20}, P_{22}, P_{24}, P_{26}, P_{28}, P_{30}$}      \\
 \hline
  \multicolumn{1}{|c||}{15}   & \multicolumn{1}{|c||}{$P_{20}, P_{23}, P_{24}, P_{27}, P_{28}, P_{31}$}      \\
 \hline
   \multicolumn{1}{|c||}{16}   & \multicolumn{1}{|c||}{$P_{21}, \cdots, P_{31}$}      \\
 \hline

\end{tabular}
\end{center}
\label{table:4}
\end{table}

\end{example}

For Polar codes with greater length, those indices of information bits which for them codeword's weights are equal to each other can be chosen. By this way, all members would have equal access set and the secret sharing would be democratic.

\section{Conclusion}

To protect a secret, secret sharing among members can be done. In secret sharing scheme, only a specified union of members can reconstruct secret. Considering Polar codes structure, a dealer can use the secret sharing based on Polar codes to send their shares in the channel which code is built on it. By using Polar codes, members can decode their shares reliably.


\begin{thebibliography}{1}

\bibitem{arikan2009channe}
Arikan, Erdal. "Channel polarization: A method for constructing capacity-achieving codes for symmetric binary-input memoryless channels." IEEE Transactions on Information Theory 55.7 (2009): 3051-3073.

\bibitem{asmuth1983modular}
Asmuth, Charles, and John Bloom. "A modular approach to key safeguarding." IEEE transactions on information theory 29.2 (1983): 208-210.

\bibitem{bertilsson1993construction}
Bertilsson, Michael, and Ingemar Ingemarsson. "A construction of practical secret sharing schemes using linear block codes." Advances in Cryptology—AUSCRYPT'92. Springer Berlin/Heidelberg, 1993.

\bibitem{blakley1979safeguarding}
Blakley, George Robert. "Safeguarding cryptographic keys." Proc. of the National Computer Conference1979 48 (1979): 313-317.

\bibitem{diffie1976new}
Diffie, Whitfield, and Martin Hellman. "New directions in cryptography." IEEE transactions on Information Theory 22.6 (1976): 644-654.

\bibitem{ding2003covering}
Ding, Cunsheng, and Jin Yuan. "Covering and secret sharing with linear codes." Discrete Mathematics and Theoretical Computer Science. Springer Berlin Heidelberg, 2003. 11-25.

\bibitem{elgamal1985public}
ElGamal, Taher. "A public key cryptosystem and a signature scheme based on discrete logarithms." IEEE transactions on information theory 31.4 (1985): 469-472.

\bibitem{gennaro1996robust}
Gennaro, Rosario, et al. "Robust threshold DSS signatures." International Conference on the Theory and Applications of Cryptographic Techniques. Springer Berlin Heidelberg, 1996.

\bibitem{kaya2014sharing}
Kaya, Kamer, and Ali Aydın Selçuk. "Sharing DSS by the Chinese remainder theorem." Journal of Computational and Applied Mathematics 259 (2014): 495-502.

\bibitem{li2015error}
Li, Liping, Wenyi Zhang, and Yanjun Hu. "On the error performance of systematic polar codes." arXiv preprint arXiv:1504.04133 (2015).

\bibitem{Liu}
Liu, Chung Laung. "Introduction to combinatorial mathematics." (1968).

\bibitem{mao2003modern}
Mao, Wenbo. Modern cryptography: theory and practice. Prentice Hall Professional Technical Reference, 2003.

\bibitem{massey1993minimal}
Massey, James L. "Minimal codewords and secret sharing." Proceedings of the 6th Joint Swedish-Russian International Workshop on Information Theory. 1993.

\bibitem{mignotte1983share}
Mignotte, Maurice. "How to share a secret." Cryptography (1983): 371-375.

\bibitem{moradi2017combining}
Moradi, Mohsen. "Combining and steganography of 3d face textures." arXiv preprint arXiv:1702.01325 (2017).

\bibitem{de1994share}
De Santis, Alfredo, et al. "How to share a function securely." Proceedings of the twenty-sixth annual ACM symposium on Theory of computing. ACM, 1994.

\bibitem{shamir1979share}
Shamir, Adi. "How to share a secret." Communications of the ACM 22.11 (1979): 612-613.





\end{thebibliography}
\end{document}